\documentclass[reprint,amsmath,amssymb,aps]{revtex4-1}
\usepackage{graphicx}
\usepackage{amssymb}
\usepackage{bm}
\begin{document}
\setcounter{page}{1}
\title[]{Embedded Diagrams of Kerr and Non-Kerr Black Holes}
\author{Yong-Jin \surname{Han}}
\email{yjhan7@sch.ac.kr}
\thanks{Fax: +82-41-530-1229}
\author{Hyun-Chul \surname{Lee}}
\affiliation{Department of Physics, Soonchunhyang University, Asan
646}

\date{\today}

\begin{abstract}
Embedded diagrams are drawn for investigating the black hole of Kerr and non-Kerr metric.
Kerr black holes are characterized by masses $M$ and spin parameters $a$. Non-Kerr black holes also are characterized by the
deformation parameters $\epsilon$, which is related to shape of the black hole, in addition to their masses and spins. In this paper, we investigate the behavior of non-Kerr black holes compared with Kerr black holes in the parameter space of ($a$,$\epsilon$) using embedded diagrams. The event horizons and the naked singularity of non-Kerr BHs are discussed in detail.

\end{abstract}

\pacs{84.40.Ik, 84.40.Fe}

\keywords{Embedded diagram, Black hole, Kerr metric, Non-Kerr metric}

\maketitle

\section{INTRODUCTION}
Space-time curvature of four-dimension is hard to conceive. Specially,
the curvature of strong gravity is too far from our common sence to percieve.
Embedded diagram method\cite{R1,R2,R3} is one of the ways to bypass these difficult.
Mathematically, embedded diagrams are two-dimensional conformal mapped surfaces embedded in three-dimensional
Euclidean space. Generally, embedded diagram represents for intrinsic geometry of the equatorial plane around black hole at constant time. The effects of parameters like mass $M$, charge $Q$ and angular momentum $J$ (or spin parameter $a = J/Mc$) of black holes are explicitly visualized in the two-dimensional curvatures of embedded diagrams and the influences are qualitatively understood very well.\cite{R2,R3}

By the no-hair theorem, neutral black holes are uniquely characterized by their masses and spins. Kerr black hole is characterized by mass $M$ and spin $a$. Thus, the metric of Kerr black hole has the stationary, axisymmetric, and asymptotically flat space-time. But, the Kerr metrics have lacked the direct experiment evidences. Several experiments of the electromagnetic wave\cite{R4,R5,R6,R7,R8} and gravitational wave\cite{R9,R10,R11,R12,R13,R14,R15} show the deviation from the Kerr metrics.

A deformed Kerr-like metric for strong gravitation in an alternative theory of general relativity has been proposed by Johannsen and Psalitis \cite{R16}.
The non-Kerr metric has the deformation parameter $\epsilon$ in addition to mass $M$ and spin $a$. They applied the Newman-Janis transformation \cite{R17} and constructed a Kerr-like metric. For positive parameter $\epsilon$, the black hole is more prolate than spherical black hole. For negative $\epsilon$, the black hole is more oblate. When the deformation parameter $\epsilon$ vanish, the metric reduces to the Kerr metric. There were many interesting properties in the non-Kerr black hole
that some researchers have reported so far. For positive $\epsilon$, there are naked singularity region for some values of $\left|a\right|/M$ \cite{R16}. For negative $\epsilon$, the event horizons always exist for any values of $\left|a\right|/M$ and the horizons are in the shape of toroidal \cite{R18, R19}. The properties of ergosphere and energetic processes of non-Kerr black holes have been investigated in the paper \cite{R20}.

In this paper, we investigate the behavior of non-Kerr black holes (BH) compared with Kerr BHs in the parameter space of ($a$,$\epsilon$) using embedded diagrams. Specially, the event horizons and the region of naked singularity of non-Kerr BHs are studied in detail.

\begin{figure}[t!]
  \begin{center}
  \includegraphics[width=\columnwidth]{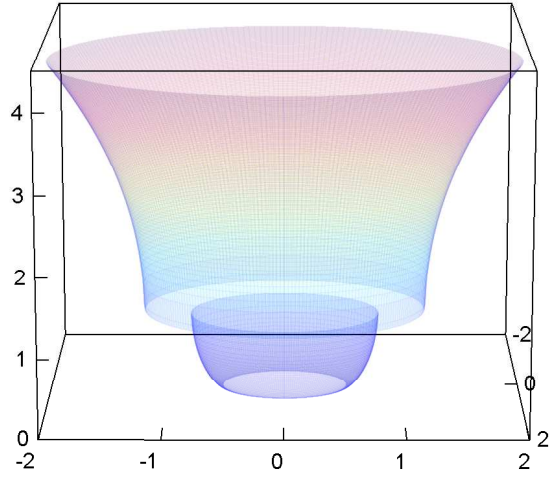}
   \caption{Embedded diagrams of Kerr BH in cylindrical coordinates ($r,\phi,z$) for spin $a=0.98$. The gap between the surfaces is
   boundary of the inner and outer event horizons.}
   \label{fig.1}
\end{center}
\end{figure}

\section{EMBEDDED DIAGRAM AND NON-KERR METRIC}

\begin{figure}[t!]
  \begin{center}
  \includegraphics[width=\columnwidth]{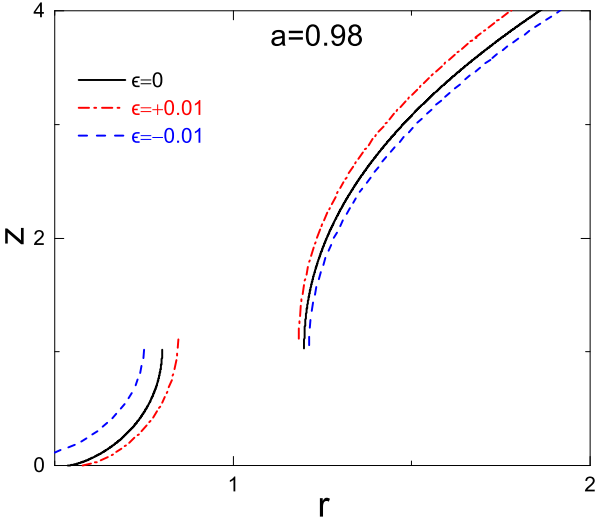}
   \caption{Embedded diagrams of Kerr BH and non-Kerr BH. Three different values of deformation parameter $\epsilon$ for the same value of spin $a=0.98$ are plotted. We can find that the small deviation($\epsilon=\pm0.01$) of Kerr BH($\epsilon=0$) makes notable differences. }
   \label{fig.2}
\end{center}
\end{figure}

\subsection{Embedded Diagram}
Static rotating axially symmetric metric is
\begin{equation}
  ds^{2}= g_{tt}dt^{2}+g_{rr}dr^{2}+g_{\theta\theta}d\theta^{2}+g_{\phi\phi}d\phi^{2}
  + 2g_{t\phi}dtd\phi.
\end{equation}
At constant time and equatorial plane ($\theta = \pi/2$), the metric becomes
\begin{equation}
  ds^{2}= g_{rr}dr^{2} + g_{\phi\phi}d\phi^{2}.
\end{equation}
We denote the embedded diagram of the metric in cylindrical coordinates as
\begin{equation}
ds^{2}=dR^{2}+R^{2}d\phi^{2}+dz^{2}.
\end{equation}
We can find a surface $z$ = $z$($R$) that is isometric to two dimensional equatorial plane ($\theta = \pi/2$). The conformal metric is thus rewritten the surface,
 \begin{equation}
dl^{2}=\Bigg[\Big(\frac{dR}{dr}\Big)^{2}+\Big(\frac{dz}{dr}\Big)^{2}\Bigg]dr^{2}+R^{2}d\phi^{2}.
\end{equation}
We require that the line element (4) should be to equal to the metric (2), therefore the following equation must be satisfied
\begin{equation}
g_{rr}=\Big(\frac{dR}{dr}\Big)^{2}+\Big(\frac{dz}{dr}\Big)^{2}, \qquad g_{\phi\phi}=R^{2}
\end{equation}
and so we constructed
\begin{equation}
\frac{dz}{dr} = \pm\sqrt{g_{rr} - \Big(\frac{dR}{dr}\Big)^{2}}.
\end{equation}
For the reason that above the expression is beyond compute by hand, we numerically calculate to find the slope that demonstrates the curvature of space-time geometry. Eq. (6) must be real number, thus so called embedded condition
\begin{equation}
g_{rr} > \Big(\frac{dR}{dr}\Big)^{2}
\end{equation}
must be satisfied.

\subsection{Non-Kerr Metric }
In the Boyer-Lindquist coordinates and in the geometric unit ($c=1, G=1$), the metric becomes \cite{R16,R20}
\begin{equation}
ds^{2} =
g_{tt}dt^{2}+g_{rr}dr^{2}+g_{\theta\theta}d\theta^{2}+g_{\phi\phi}d\phi^{2}+2g_{t\phi}dtd\phi,
\end{equation}
with
\begin{equation}\label{eq1}
\begin{split}
g_{tt} = -\Big(1-\frac{2Mr}{\rho^{2}}\Big)(1+h),
g_{t\phi} = -\frac{2aMr\sin^{2}\theta}{\rho^{2}}(1+h),\\
\\
g_{rr} = \frac{\rho(1+h)}{\triangle+a^{2}h\sin^{2}\theta}, \quad
g_{\theta\theta} = \rho^{2},\\
\\
g_{\phi\phi} =
\sin^{2}\theta\Bigg[\rho^{2}+\frac{a^2(\rho^{2}+2Mr)\sin^2\theta}{\rho^{2}}(1+h)\Bigg],
\\
\end{split}
\end{equation}
where
\begin{equation}
\rho^2 = r^2+a^2\cos^2\theta, \quad \triangle = r^2-2Mr+a^2, \quad h = \frac{\epsilon M^3
r}{\rho^4}.
\end{equation}
\\
The metric has the deformation parameter $\epsilon$ that represents the degree of variation that
the black hole is more prolate ($\epsilon$ $>$ 0) or oblate($\epsilon$ $<$ 0) than the
kerr black hole. Kerr metric is restored when $\epsilon$ is zero.\\

The event horizon of a black hole is described by \cite{R3}
\begin{equation}
\Delta+ a^2 h \sin^{2}\theta = 0.
\end{equation}
The radii of event horizons of non-Kerr black hole depend on $\theta$.

\begin{figure}[t!]
  \begin{center}
  \includegraphics[width=1\columnwidth]{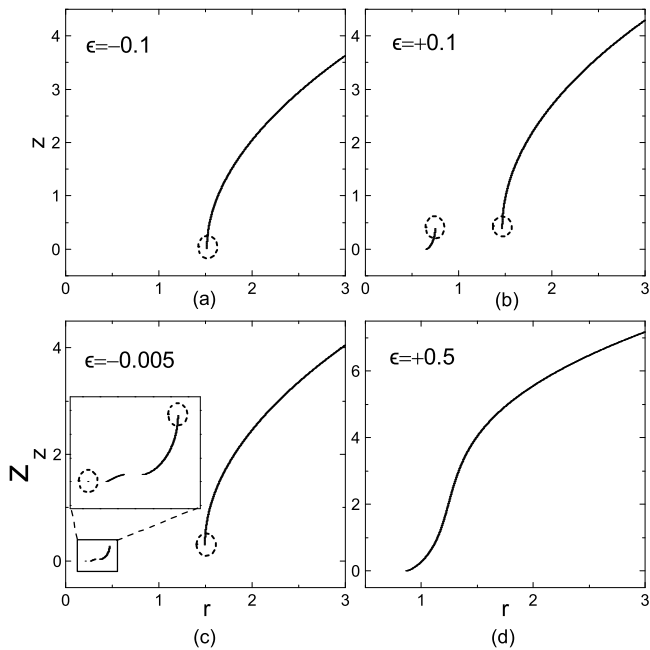}
   \caption{Four different phases in embedded diagrams of the non-Kerr BH in parameter space ($a, \epsilon$) ; (a) the phase of single event horizon (EH), (b) two EHs, (c) three EHs, and (d) no EH (naked singularity). EHs are noted by the small broken-lined circles. Spin $a=0.87$ for all plots. }
   \label{fig.3}
\end{center}
\end{figure}

\section{Non-Kerr Black Holes in Parameter Space}
\subsection{Comparison Kerr BH with non-Kerr BH}
We use $M=1$ from now on. We draw the embedded diagrams at constant time and equatorial plane ($\theta = \pi/2$). There are many interesting features in non-Kerr BH compared with Kerr BHs. Even the small value of deformation parameter $\epsilon$ makes notable difference in behaviors of BH. We can find the aspects in the embedded diagram in Fig. 1 and 2.
Kerr BHs has two event horizons (inner and outer event horizon) for $a<1$ and naked singularity for $a>1$.
In general, there are four different phases in embedded diagrams of the non-Kerr BH in parameter space ($a, \epsilon$); the phase of single event horizon (EH), two EHs, three EHs, and no EH(naked singularity)(see Fig 3). We will discuss these in detail in the section C.

The difference $\Delta r$ of inner event horizon $r_{-}$ and outer horizon $r_{+}$ is plotted in Fig. 4 ($\Delta r$ vs. $a$). In case of Kerr BHs, $\Delta r$ is propotional to $\sqrt{1-a^{2}}$. However, in case of non-Kerr BHs, it's dependence is deviated from that as shown in Fig. 4.

In case of Kerr BH, there is no naked singularity in $a < 1$. In case of non-Kerr BH, there are naked singularities for any value of $a$ . Following section, we will discuss about naked singularity of non-Kerr BH in detail.

\begin{figure}[t!]
  \begin{center}
  \includegraphics[width=\columnwidth]{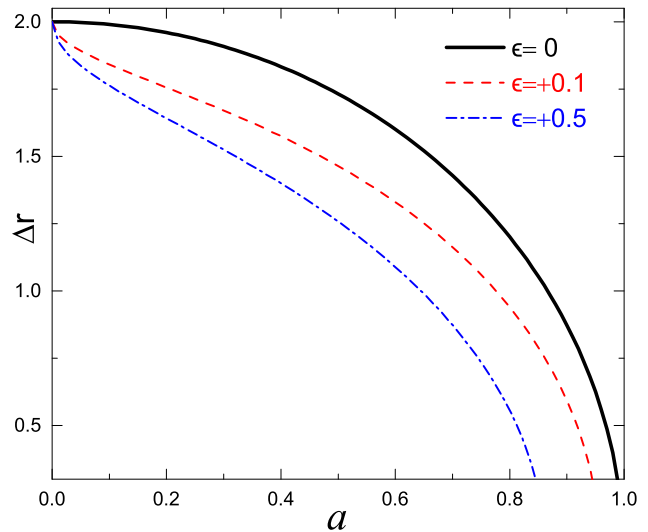}
   \caption{The difference $\Delta r$ of inner event horizon $r_{-}$ and outer horizon $r_{+}$ for Kerr BH and non-Kerr BH. In case of Kerr BHs (black thick line), $\Delta r$ is propotional to $\sqrt{1-a^{2}}$. However, in case of non-Kerr BHs, it's dependence is deviated from that as shown in figure.}
   \label{fig.4}
\end{center}
\end{figure}

\begin{figure}[t!]
  \begin{center}
  \includegraphics[width=\columnwidth]{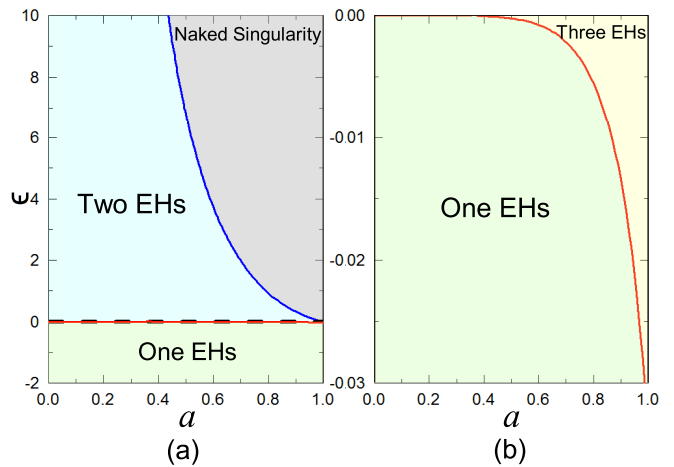}
   \caption{Four phases of EH of non-Kerr BHs; the phase of single EH, two (inner and outer) EHs, three EHs, and no EH (naked singularity). Four phases is displayed in the parameter space ($a, \epsilon$). The region of three EHs is very narrow and small ($0>\epsilon> -10^{-2}$) in values of parameter $\epsilon$.}
   \label{fig.5}
\end{center}
\end{figure}

\subsection{Naked Singularity of non-Kerr BH}
In case of Kerr BH, the condition for the existence of the naked singularity is only $a>1$.
The condition for the existence of the naked singularity for non-Kerr BH is $\epsilon \geq 0$ \cite{R16,R19}. There is no naked singularity in negative $\epsilon$. There is no restriction of value of spin $a$ for having naked singularity. In Fig.5(a), for the region of naked singularity, one can see $a\rightarrow 1^{-}$ as $\epsilon \rightarrow 0^{+}$. Less prolate BH is, faster spin $a$ is \cite{R19}. But, in nature, the possibility of existence of BH with $a \simeq 1$ is not high because of its unphysically high angular momentum. Non-Kerr BHs with $a<1$ having naked singularity may be existed in nature. But, in case of $\epsilon > 0$, the BH is prolate one that also is unlikely created in nature because of its high centrifugal force. Thus, we conclude that, at least in case of non-Kerr BH, there is no possibility for existence of naked singularity BH. In this context, the assumption of the Weak Cosmic Censorship Conjecture (WCCC) could be satisfied.

In embedded diagram, the slope ($dz/dr$) of starting positions of naked singularity for Kerr or non-Kerr BHs are almost zero, i.e. ($dz/dr \simeq 0$) while the slope of BHs at the event horizons are infinity ($dz/dr = \infty$) at their starting points (see the Fig. 3).

\subsection{Event Horizons of non-Kerr BH}

The variety of non-Kerr BHs emerges in aspect of event horizon.
There are four phases for EHs of non-Kerr BHs; the phase of single EH, two (inner and outer) EHs, three EHs, and no EH (naked singularity. The four phases are described in Fig. 5.

In case of single EH, the non-Kerr BH acts like Schwarzschild BH.
In case of two EHs, the non-Kerr BH acts like Kerr BH.
In case of three EHs, an event horizon (EH) in addition to inner and outer horizons exists in special region of parameter space ($a$,$\epsilon$). The region is very narrow and small ($0>\epsilon> -10^{-2}$) in values of parameter $\epsilon$ shown in Fig. 5b.
The third EH is deep in the inner EH of the non-Kerr BH as shown in Fig. 6.  We can see from Fig. 6 that EH points are located at $g_{rr} = \infty$ and there are three EHs. In this case, embedded diagram shows the slope of $dz/dr$ is always infinity at the point of EH.
But, for even small $|\epsilon|$ ($0>\epsilon> - 10^{-3}$), the third EH is not shown in the embedded diagram because the embedded condition Eq.(7) is not satisfied .

\begin{figure}[t!]
  \begin{center}
  \includegraphics[width=\columnwidth]{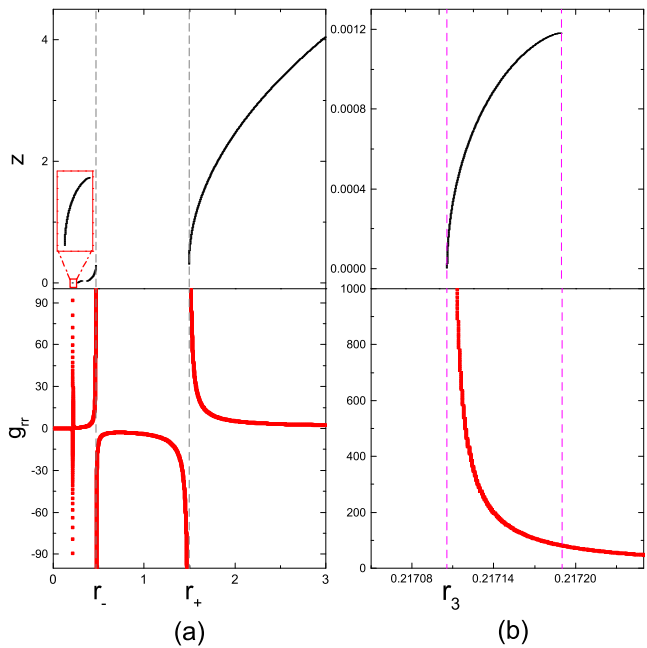}
  \caption{The third EH of deep in the BH. This figure is the case of Fig. 3(c). (b) shows the enlargement of inset of (a). We can see that EH points are located at $g_{rr} = \infty$ and there are three EHs (inner EH $r_{-}$, outer EH $r^{+}$ and third EH $r_{3}$).  In this case, embedded diagram shows the slope of $dz/dr$ is infinity at the point of EH.  }
   \label{fig.6}
\end{center}
\end{figure}


\section{CONCLUSIONS}
We have investigated event horizons(EHs) and naked singularities of non-Kerr BHs compared with Kerr BHs.
Kerr BHs have not naked singularities for $a<1$.
Non-Kerr BH have naked singularities for $a<1$ because of deformation parameter $\epsilon$.
However, naked singularity of non-Kerr BHs with $a<1$ and $\epsilon > 0$ may not be existed in nature because the BH with $\epsilon > 0$ is prolate one that is unlikely created in nature. Thus, in this context, the assumption of the Weak Cosmic Censorship Conjecture (WCCC) could be satisfied.

There are four phases for EHs of non-Kerr BHs; the phase of single EH, two (inner and outer) EHs, three EHs, and no EH (naked singularity).
In case of single EH, the non-Kerr BH acts like Schwarzschild BH.
In case of two EHs, the non-Kerr BH acts like Kerr BH.
In case of three EHs, an event horizon in addition to inner and outer horizons exists.
This third EH is new aspect of BH physics.

We have investigated only in case of $\theta = \pi/2$. The investigation for arbitrary $\theta$ might provide more variety of non-Kerr BH \cite{R19,R20}.



\end{document}